\documentclass[review]{elsarticle}

\usepackage{hyperref}

\usepackage{graphicx}
\usepackage{xcolor}

\journal{Astronomy and Computing}

\biboptions{authoryear}

\begin{document}

\begin{frontmatter}

\title{Low Mass X-ray Binary Simulation Data Release}%\tnoteref{mytitlenote}}
%\tnotetext[mytitlenote]{Fully documented templates are available in the elsarticle package on \href{http://www.ctan.org/tex-archive/macros/latex/contrib/elsarticle}{CTAN}.}

%% Group authors per affiliation:
%\author{Chatrik Singh Mangat\fnref{myfootnote}}
%\address{BITS Hyderabad}
%\author{Natalia Ivanova\fnref{myfootnote2}}
%\address{UAlberta}
\fntext[copyright]{© 2022. This manuscript version is made available under the CC-BY-NC-ND 4.0 license https://creativecommons.org/licenses/by-nc-nd/4.0/}
%\fntext[myfootnote2]{What2}

%% or include affiliations in footnotes:
\author[address1,address2]{Chatrik Singh Mangat\corref{mycorrespondingauthor}}
\ead{chatrikmangat@outlook.com}
\cortext[mycorrespondingauthor]{Corresponding author}

\author[address2]{Natalia Ivanova}
\ead{nata.ivanova@ualberta.ca}

\author[address2,address3]{Kenny Van}

\address[address1]{Department of Physics, Birla Institute of Technology and Science, Hyderabad, India}
\address[address2]{Department of Physics, University of Alberta, Canada}
\address[address3]{WCB Jarvis Building, Edmonton AB, Canada}

\begin{abstract}
We present {\tt DABS} (Database  of Accreting Binary Simulations), an open-access database of modelled Low Mass X-ray Binaries (LMXBs).  {\tt DABS} has been created using evolutionary tracks of neutron star and black hole LMXBs, spanning a large set of initial conditions for the accretor mass, donor mass, and orbital period. The LMXBs are evolved with the Convection and Rotation Boosted Magnetic Braking prescription.
The most important asset of this online database is the tool {\tt PEAS} (Progenitor Extractor for Accreting Systems). This tool can be used to predict the progenitors of any user-entered LMXB system and view their properties before the start of mass transfer. This prediction can facilitate preliminary searches for the progenitors of observed  LMXBs, which can help in streamlining further detailed analyses.
The {\tt PEAS} tool can also be used to constrain population synthesis techniques that specialize in supernova kicks in binaries and common envelope outcomes.
\end{abstract}


\end{frontmatter}

\section{Introduction}
\label{sec:intro}

An X-ray binary (XRB) is a system containing a donor star and an actively accreting compact object. XRBs are some of the most well-studied binary systems in astronomy because they can provide crucial information for understanding the underlying neutron star (NS) and black hole (BH) populations. Whilst millisecond pulsars provide a robust way for gaining insight into the NS population \citep[e.g., see the overview of methods in][]{2021arXiv210110081V}, it is much more challenging to observe stellar-mass BHs without active accretion. New methods to observe compact objects in non-interacting binaries using the orbital motion of the visible star or radial velocity measurements, however, show promising results for relating population synthesis outcomes to large-scale survey expectations  \citep[e.g., see][]{Weller2022,Wiktor2020}. 

Importantly, XRBs represent binaries that have survived various transformative phases of stellar evolution, like a common envelope (CE) event or the supernova of the high-mass companion. As a result, XRB populations can represent more direct progenitors for compact binaries observed by gravitational wave observatories than the underlying stellar binary populations. However, even with the reduced complexity of transformations, comparing the XRB population to that of compact binaries is not straightforward, as different formation channels and selection effects play an important role in the process \citep{Fishbach2021}. 

The evolution of XRBs is driven by the individual stars' evolution and by binary processes like mass transfer (MT) and orbital angular momentum loss. The MT exhibited by these systems can occur through two main processes: wind-fed accretion in High Mass X-ray Binaries (HMXBs) and Roche Lobe Over-Flow (RLOF) in Low Mass X-ray Binaries (LMXBs). The orbital angular momentum can be lost due to mass loss from the binary system, through gravitational waves \citep{Faulkner1971} when the orbital separation is low, and due to magnetic braking (MB) \citep{Rappaport1983}. The latter mainly operates when a tidally locked donor has a convective envelope. Recent studies have shown that the widely used Skumanich prescription for MB cannot reproduce MT rates in the NS LMXBs that are observed to be persistently bright \citep{Van2019}. Instead, the observed persistent NS LMXBs can be reproduced using the Convection And Rotation Boosted (CARB) MB prescription  \citep{Van2021}. 

Currently, there is a well-known disconnect between the needs of observational studies of individual LMXBs of our galaxy, and the interests of theoretical studies. Usually, theoretical studies are either devoted to individual well-known LMXBs presented along with a new (or modified) theoretical tool, or to statistical formation studies of LMXB populations. We want to bridge the gap between the observational and theoretical studies of LMXBs, with the goal that observers would be able to predict likely progenitors of the LMXBs that they observe, without invoking additional theoretical studies. This can be done when the progenitor population can be directly related to observationally derived LMXB properties such as mass ratio, orbital period, MT rate and the donor's effective temperature.

Hence, we present {\tt DABS} (Database of Accreting Binaries), an open-access database of LMXB simulations with NS and BH accretors using the CARB MB prescription. This database contains detailed evolutionary tracks of LMXBs, which can be directly used for future population studies. Moreover, we present {\tt PEAS} (Progenitor Extractor for Accreting Systems), a tool that uses our simulation data to predict viable progenitors of any LMXB system with user-entered properties. The NS database contains the simulation data described in \cite{Van2021}, and we perform similar simulations for the BH database with a few modifications (See Section. \ref{sec:sim}). In this work, progenitor systems are considered to be binary systems that have survived poorly-constrained phases such as the supernova kick and CE events. {\tt DABS} is intended to inform the choices of populations in comparison studies and constrain the progenitors of any observed LMXB. 
Future measurements of observed LMXB properties will provide better constraints for progenitor prediction using the database, it may also allow constraining disk instability models. Moreover, the restriction of the progenitor space will hopefully help us further constrain CE outcomes and natal kicks. Extensive author-driven analyses of progenitor systems of NS LMXBs have been published in \citet{Van2021}. The creation of the presented open-access database, however, was prompted by requests for LMXB progenitor searches with different constraints.

This paper is arranged as follows: in Section~\ref{sec:sim}, we outline the simulation setup and the differences between the NS and BH databases. In Section~\ref{sec:tab}, we describe the progenitor grid used for the simulations, along with the working of the progenitor prediction tool. Section~\ref{sec:concl} concludes our work and describes avenues for future research that our database can facilitate.

\section{{\tt{MESA}} Simulations}

\label{sec:sim}

We use the 1-D stellar code {\tt{MESA}} \citep{Paxton2011,Paxton2013,Paxton2015,Paxton2018,Paxton2019} to simulate our LMXB systems.  The code offers a multitude of options to define the physical processes that we want to use for our simulations. Throughout this work, we use the {\tt{MESA}} version 15140 unless specified otherwise.

The NS LMXBs in our database have been simulated previously, and details about the simulation setup for MT can be found in \citet{Van2021}. For BH LMXBs, we adopted the default treatment of the Eddington limit for BH accretors implemented in {\tt MESA} \citep{Paxton2015}, which was originally described in \citet{Podsiadlowski2003}.
The apparent luminosity due to accretion is given by:

\begin{equation}
L_{\rm Acc}=\frac{\epsilon \dot{M} c^{2}}{f} \ ,
\end{equation}

\noindent where $\dot{M}$ is the accretion rate, $\epsilon$ is the accretion efficiency and $f=6$ for a non-rotating BH, 1 for a maximally rotating BH with co-rotating orbits, and 9 for retrograde orbits. This gives the Eddington limit for accretion onto a BH:  

\begin{equation}
\dot{M}_{\rm Edd}=\frac{f}{\epsilon}\frac{4 \pi G M}{ c \kappa_{e}} \approx \frac{f}{\epsilon} \frac{4.6 \times 10^{-9}}{1+X}
\frac{M}{M_\odot} M_\odot {\rm yr}^{-1} \ .
\end{equation}

\noindent Here $\kappa_{e}$ is the opacity due to Thomson electron scattering. In our case, we define $\kappa_{e}=0.2(X+1)$ $\rm g^{-1} cm^{2}$ where $X$ is the mass fraction of hydrogen in the envelope of the donor. Additionally, we assume non-conservative MT with an efficiency $\eta = 0.5$, which implies that only half the material removed from the donor during RLOF is accreted, while the rest is lost from the vicinity of the accretor as a fast wind. Thus, the upper limit for MT is derived from the Eddington limit and the accretion efficiency:

\begin{equation}
\dot{M} = {\rm min} \{ \eta \dot{M}_{\rm tr} , \dot{M}_{\rm Edd}\} \ ,
\end{equation}

\noindent where $\dot{M}_{\rm tr}$ is the mass lost by the donor due to RLOF.

The details related to the treatment of single star evolution which {\tt{MESA}} requires (winds, convection treatment, nuclear networks, photospheric boundary condition and precision)  are provided in the inlists enclosed with the database. Additionally, we provide information regarding the subroutines used to model the MT rates as described above, for the NS and BH simulations \href{https://github.com/ChatrikMangat/progenlmxb}{here}\footnote{GitHub Repository: https://github.com/ChatrikMangat/progenlmxb}.

\section{Database \& Progenitor Prediction}
\label{sec:tab}

The database is divided into four sets based on the accretor used in the simulations (1.4 $M_\odot$ NS or 5, 7, 10 $M_\odot$ BH \footnote{Additional sets with BH of different masses can be added to the database in the future if there will be demand.}). For each accretor, we construct a grid of initial conditions for the simulations, based on the donor mass and orbital period. We divide each grid into four sections based on the mesh resolution for these two properties (similar to \cite{Van2021}) as shown in Table \ref{tab:init}. For each initial configuration, we evolve a binary system in MESA using the controls specified in Section~\ref{sec:sim}. The results of the simulation are trimmed and we store the time evolution of selected binary properties in our database, namely: donor mass ($M_\odot$), accretor mass ($M_\odot$), $\rm log_{10} (MT \: rate/(M_\odot/{year})$), orbital period (days), $\rm log_{10} (donor \; T_{eff}/(K))$, age of system (years), $\rm log_{10} (donor \: radius/R_\odot)$ and $\rm log_{10} (timestep/years)$. Please note that the age of system represents the time elapsed since the start of the simulation, not the actual age of the component stars.  This trimmed database, which is also used by the {\tt PEAS} tool, can be accessed through Zenodo \citep{ZenodoDB}. We also provide scripts to create the pre-set directory structure and all input files for our {\tt MESA} simulations in our \href{https://github.com/ChatrikMangat/progenlmxb}{repository}, in case there is a need to repeat specific simulations locally. Moreover, the raw {\tt{MESA}} output, which keeps track of many additional binary properties for all our simulations, is available on request\footnote{{The size of the raw data is 9.3 TB.}}.

\begin{table}[ht]
    \centering
    \begin{tabular}{ccccc}
        \hline
         Tag & $M_d$ range & $\Delta M_d$ & $\log_{10}{(P)}$ range & $\Delta \log_{10}{(P)}$ \\
         \hline
        SMSP & 0.95 $-$ 4.00 & 0.05 & -0.60 $-$ 1.64 & 0.02 \\
        SMLP & 0.95 $-$ 4.00 & 0.05 &  1.65 $-$ 4.00 & 0.05 \\
        LMSP & 4.00 $-$ 7.00 & 0.10 & -0.60 $-$ 1.64 & 0.02 \\
        LMLP & 4.00 $-$ 7.00 & 0.10 &  1.65 $-$ 4.00 & 0.05 \\
        \hline
    \end{tabular}
    \caption{The grid ranges and intervals for the initial conditions of the simulations. Here, $M_d$ is the mass of the donor ($M_\odot$) and $P$ is the orbital period of the binary (days).}
    \label{tab:init}
\end{table}

We now describe the algorithm we use to find progenitors of a given LMXB system. To check if a simulated system is a possible progenitor of an observed system, we simultaneously match its donor mass, accretor mass, orbital period, MT rate and donor $T_{\rm eff}$ (if provided by the user) with the observed system. If a match is found at any time in the evolution of the simulated system, it is considered to be a progenitor of the observed system. This principle is implemented in the {\tt PEAS} tool. A user can enter the desired range of the properties to be matched and send it as a query to the tool. The tool then searches through the database and outputs all possible progenitors of the queried LMXB system. 

We provide query format guidelines and samples with the python-based tool in our \href{https://github.com/ChatrikMangat/progenlmxb}{repository}. In the query, the user can specify whether the tool should search through the NS simulation set or the BH simulation sets. If the user wants to search through the BH sets, we select the relevant BH sets by checking the initial BH mass of the simulation. A set is skipped if the initial BH mass is more than the upper BH mass limit of the query or if it is less than (lower BH mass limit of the query - $\Delta M_{\rm BH}^{\rm max}$). We use $\Delta M_{\rm BH}^{\rm max} = 3.5 M_\odot$, which is the maximum mass that can be accreted by a BH in our simulation setup. After the sets have been selected, the following pseudo-algorithm describes how the tool searches for matches in one simulation data file with $N$ lines:

\begin{enumerate}
    \item Check the initial donor mass of the simulation:
    \begin{enumerate}
        \item If it is less than the lower donor mass limit of the query, skip the file.
        \item Else, continue.
    \end{enumerate}
    \item Check the final donor mass of the simulation:
    \begin{enumerate}
        \item If it is higher than the upper donor mass limit of the query, skip the file.
        \item Else, continue.
    \end{enumerate}
    \item Check the final accretor mass of the simulation:
    \begin{enumerate}
        \item If it is less than the lower accretor mass limit of the query, skip the file.
        \item Else, define variables $startm$ and $endm$ to signify the positions of upper and lower donor mass limits of the query, respectively, in the data file.
    \end{enumerate}
    \item Check the initial donor mass of the simulation:
    \begin{enumerate}
        \item If it is less than the upper donor mass limit of the query, set $startm = 0$.
        \item Else, do a binary search to find the location of upper donor mass limit in the data and initialize $startm$.
    \end{enumerate}
    \item Check the final donor mass of the simulation:
    \begin{enumerate}
        \item If it is more than the lower donor mass limit of the query, set $endm = N-1$.
        \item Else, do a binary search to find the location of lower donor mass limit in the data and initialize $endm$.
    \end{enumerate}
    \item Check if $[m = endm - startm] > 0$:
    \begin{enumerate}
        \item If true, do a linear search from $startm$ to $endm$ in the file and match all remaining properties of each simulated model against the query. If a match is found, calculate relevant information for output and add the system to the list of progenitors.
        \item Else, skip file.
    \end{enumerate}
\end{enumerate}

This process is repeated for every simulation data file in the database, effectively searching through relevant simulation data to find progenitors for the queried system. The processing time of the tool depends on the input given by the user. Tight constraints on the observed properties will limit the search space and increase the tool's efficiency. Most importantly, a tight constraint on the donor mass will guarantee efficient search times. It is also possible to search for the progenitors of a system without constraints on certain properties. This should not hinder the efficiency of the tool as long as the constraint on the donor mass is well defined. 

\begin{figure}[!ht]
    \includegraphics[width=1.0\textwidth]{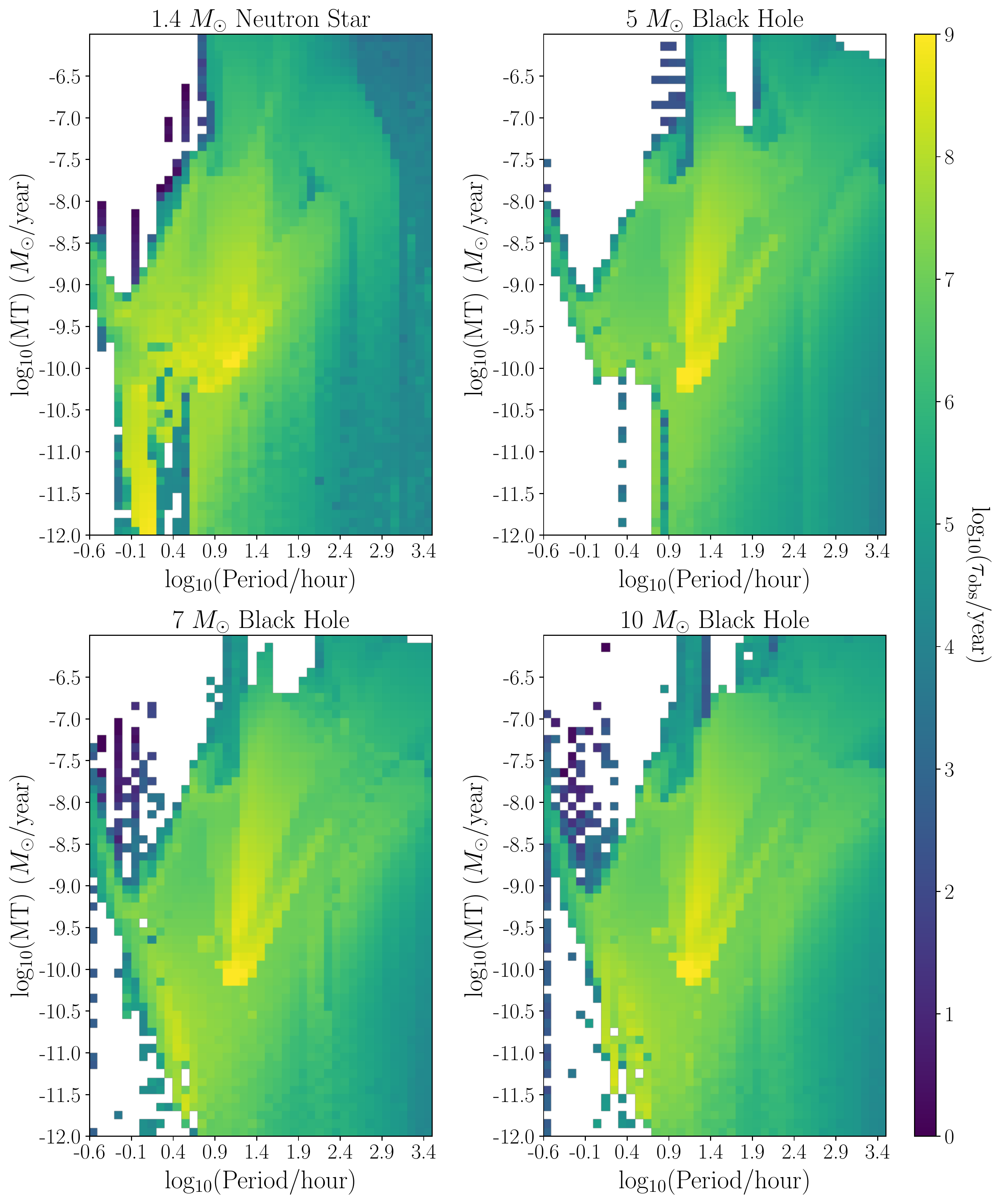}
    \caption{The maximum time ($\tau_{\rm obs}$) spent in each MT-Period bin by any simulated system. Presented for all four simulation sets based on the choice of accretor. White space indicates that no simulated system spent more than a year in that MT-Period bin.}
\label{Fig:maxtime}
\end{figure}

A preliminary analysis of whether a progenitor for a system with a given MT rate and orbital period can be found can be made using Figure \ref{Fig:maxtime}. A more refined progenitor space can then be found by restricting the search further with the donor mass, accretor mass, or effective temperature.

After the search is complete, the tool will output eleven quantities for each progenitor found: the initial donor mass ($M_\odot$), $\rm log_{10} (initial \; orbital \; period/days)$, initial accretor mass ($M_\odot$), amount of time spent as the observed system ($\tau_{\rm obs}$) (years), total evolution time spanned by the simulation ($\tau_{\rm tot}$) (years), donor mass at start of MT ($M_\odot$), $\rm log_{10} (orbital \: period \: at \: start \: of \: MT/days)$, accretor mass when the simulation first matches the query ($M_\odot$), accretor mass when the simulation last matches the query ($M_\odot$), age of the system when it first matches query (years) and age of the system when it last matches query (years). Please note that the age of the system represents the time elapsed since the start of the simulation, not the actual age of the component stars. Additionally, the {\tt PEAS} tool can also be accessed directly using \href{https://t.me/astropeas}{PEAS telegram channel} \footnote{PEAS Telegram Channel: https://t.me/astropeas} for light searches and quick look-ups without downloading the data and python scripts.

\section{Conclusion}
\label{sec:concl}

We have presented {\tt DABS}, an open-access database of LMXB simulations performed with {\tt{MESA}} with a range of initial conditions for the accretor mass, donor mass and the orbital period. This database contains one set of simulations for binaries with an NS accretor and three sets of simulations with different initial BH  masses (5$M_\odot$, 7$M_\odot$ and 10$M_\odot$). It is available for download on Zenodo \citep{ZenodoDB}. We also provide the {\tt PEAS} tool, which predicts viable progenitors of any user-entered LMXB system using {\tt DABS} \href{https://github.com/ChatrikMangat/progenlmxb}{here}. This tool is also available for direct use on \href{https://t.me/astropeas}{Telegram}   to facilitate quick look-ups and searches without downloading the entire database locally. To obtain further details of the investigated observed system, users can perform their own detailed searches within the progenitor space constrained by our tool using the reference inlists and subroutines for {\tt {MESA}} provided in our GitHub \href{https://github.com/ChatrikMangat/progenlmxb}{repository}.

We hope that the presented open-access theoretical database will be helpful for observers who want to get information about the origin of the LMXBs they observe and for theorists who wish to constrain their binary population synthesis studies. A possible extension for population synthesis studies is to combine the observed systems, our progenitors, and population study codes for CE events and supernova kicks. Our database can also help streamline higher resolution searches and more detailed analyses by highlighting the most relevant sections of the progenitor space. Since processes in stellar evolution are highly non-linear, our database can facilitate more informed hypotheses about the past evolution of observed XRBs. Moreover, we hope that our tool's constraints on the progenitor space can help establish robust links between post-CE binaries and observed XRBs. 
For future work, we plan to extend this database to include more initial accretor masses to cover the progenitor parameter space more extensively.

\section{Acknowledgments}

C.S.M. thanks the Department of Physics and the Department of Computer Science and Information Systems at BITS Hyderabad for organizing two off-campus semesters. C.S.M also acknowledges the support provided by IPCD BITS Hyderabad that allowed him to pursue this work in-person at the University of Alberta. 
N.I. acknowledges funding from NSERC Discovery under Grant No. NSERC RGPIN-2019-04277. 
This research was enabled in part by support provided by Compute Canada (\url{www.computecanada.ca}). The presented database was created using about 600 CPU core years at Compute Canada.

\appendix
\section{Notes on Query Construction}

In this appendix, we show how the query input for {\tt{PEAS}} can be constructed and modified to get different progenitors populations (e.g. persistent or transient). A query consists of upper and lower bounds on a maximum of five LMXB properties as shown subsequently. Once constructed, the query can then be sent as input to the {\tt{PEAS}} tool to find the progenitors of the observed LMXB system either directly through Telegram or by downloading the database and the command line version of the {\tt{PEAS}} tool.

\subsection{A persistent system}

For this example, we use the case of the persistent NS LMXB system named Sco X-1. The query consists of upper and lower bounds for the donor mass, accretor mass, orbital period, $\rm \log_{10}{(MT \; rate)}$ and $ \rm \log_{10}{(donor \; T_{\rm{eff}})}$. The bounds used for each of these properties in our example can be seen in Table \ref{tab:obs}. The progenitor distribution for this system has been analysed previously in \citet{Van2021}, and our search algorithm yields similar results to those obtained in that work. We find a total of 39 progenitor systems for Sco X-1, with initial donor masses ranging from 1.0 $M_\odot$ to 1.9 $M_\odot$ and an initial accretor mass of 1.4 $M_\odot$. All the progenitors evolved into models that matched the observed system for a duration ($\tau_{\rm obs}$) of roughly $5*10^6$ years, and the total mass accreted by the NS during this time was around 0.02 $M_\odot$. For our case of Sco X-1, $\tau_{\rm obs}$ was mostly constrained by the effective temperature bounds on the donor star. 

Since Sco X-1 is categorised as a persistent system, we directly used MT rate estimates from \citet{Van2021} to guarantee that all progenitors predicted by our tool would result in persistent models. However, it is possible for users to derive their own bounds from observed LMXB properties. For instance, one can use the luminosity estimates provided in \citet{Cherepashchuk2021} to calculate the MT rate with their preferred model. These bounds can then be entered into the query to modify the progenitor distribution to their preferences.

\begin{table}[ht]
    \centering
    \begin{tabular}{ccc}
        \hline
         Properties & Sco X-1$_{(\rm a)}$ & XTE J1118+480$_{(\rm b)}$ \\
         \hline
         Donor Mass ($M_\odot$) & 0.30 $-$ 0.50 & 0.09 $-$ 0.50 \\
         Accretor Mass ($M_\odot$) & 1.40 $-$ 1.73 & 6.25 $-$ 7.49 \\
         Orbital Period (days) & 0.758 $-$ 0.851 & 0.166 $-$ 0.175 \\
         $\rm \log_{10}{(MT \; rate)}$ ($M_\odot$/yr)  & -7.8 $-$ -7.1 & -11.6 $-$ -10.6\\
         $ \rm \log_{10}{(Donor \; T_{\rm{eff}})}$ (K) & 3.40 $-$ 3.68 & 3.30 $-$ 4.00 \\
        \hline
    \end{tabular}
    \caption{(a) The upper and lower bounds for each queried property for Sco X-1. The mass bounds and effective temperature bounds are set according to estimates in \citet{Cherepashchuk2021} and \citet{Van2021}, and the orbital period and MT rate bounds are taken from \citet{Van2021}. (b) The upper and lower bounds for each queried property for XTE J1118+480. The mass bounds are taken directly from estimates in \citet{Chatterjee2019} and the period bounds are centered on the 4.1 hr observed period of the LMXB. The MT bounds are centered on estimates taken from \citet{Watchdog2016}. There is no estimate for the donor's effective temperature, so the bounds are kept wide enough to include all physical values.}
    \label{tab:obs}
\end{table}

\subsection{A transient system}

For this example, we use the case of the transient BH LMXB named XTE J1118+480. The observed properties available to constrain the progenitor grid for this system are donor mass, accretor mass, orbital period and $\rm \log_{10}{(MT \; rate)}$. These bounds for our example can be seen in Table \ref{tab:obs}. For this system, due to the tightness of the bounds, we find two progenitors for our observed system. Both progenitors have an initial accretor mass of 7 $M_\odot$. The progenitor systems have initial donor masses of 1.10 $M_\odot$ and 1.15 $M_\odot$ and initial orbital periods of 1.9 days and 1.4 days, respectively. The output shows that the models matched the observed system for $10^8$ years, roughly 1\% of their lifetime, and the total mass accreted by the BH during this time was less than $10^{-3} M_\odot$. 

In our example of the transient BH LMXB, the time-averaged MT rate was known, but that is not the case for most of the transient systems. We can also construct a query to ensure that all progenitors (for the specific system) predicted by the tool are transient in nature. For our example, we consider the disk instability model described in \citet{Coriat2012}, which gives us the critical MT rate for persistent MT as follows:
$$
\dot{M}_{\rm crit} = kP_{\rm hr}^b {\rm \;\; g\; s^{-1}} \ ,
$$
where $P_{\rm hr}$ is the orbital period of the LMXB system in hours. In the irradiated regime, $b=1.59$ and $k=(3.9 \pm 1.6) * 10^{15}$ for a BH accretor. All systems with MT rates below $\dot{M}_{\rm crit}$ will be classified as transient, so we can change the upper bound in our query to be the critical MT rate for XTE J1118+480, as described above. A lower bound of $10^{-14} M_\odot$/yr would ensure that all mass transferring systems are included. This wide range of MT rates will ensure that the tool provides all the progenitors that can make a transient system with the given donor mass, period, and effective temperature, while having the unknown apriori relation between the time-averaged MT rate and the outburst luminosity. This can provide some constraints on how to link the (modeled) time-averaged MT rate and the observed luminosity during the outburst.

\bibliography{dabs.bib}

\end{document}